\documentclass[11pt]{article}

\newcounter{taskcounter}

\newcounter{facetcounter}

\usepackage{multirow}
\usepackage{array}
\usepackage{ifthen}
\usepackage{amsmath}
\usepackage{amsfonts}
\usepackage{amsthm}
\usepackage{color}
\usepackage{verbatim}
\usepackage{relsize}
\usepackage{array}
\usepackage{listings}
\usepackage{rotating}
\usepackage{acronym}
\usepackage{wrapfig}
\usepackage[export]{adjustbox}

\usepackage{hyperref}
\usepackage{tikz}
\usepackage{pgfplots}
\usetikzlibrary{matrix,arrows,decorations.pathmorphing,patterns,positioning}
\usepackage{stmaryrd}
\usepackage[super,move]{cite}
\usepackage{wrapfig}
\usepackage{enumitem}
\usepackage{xcolor, soul}
\usepackage{suffix}

\usepackage{perpage}
\MakePerPage{footnote}

\usepackage{wrapfig}

\setlist[itemize]{leftmargin=0.5cm,itemsep=0cm,topsep=0.2cm}

\usepackage{titlesec}
\titlespacing*{\section}{0pt}{0.25\baselineskip}{0.25\baselineskip}
\titlespacing*{\subsection}{0pt}{0.25\baselineskip}{0.25\baselineskip}
\titlespacing*{\subsubsection}{0pt}{0.25\baselineskip}{0.25\baselineskip}
\titlespacing*{\paragraph}{0pt}{0.5\baselineskip}{0.25\baselineskip}
\titleformat{\subsubsection}
{\normalfont\normalsize\bfseries\itshape}{\thesubsubsection:}{1em}{}

\makeatletter
\@ifclassloaded{article}{
  \def\lastoftwo@dotted#1.#2\@nil{#2}
  \def\lastofthree@dotted#1.#2.#3\@nil{#3}
  \let\submynthofm\lastoftwo@dotted
  \let\subsubmynthofm\lastofthree@dotted
}{}
\@ifclassloaded{IEEEtran}{
  \def\lastoftwo@hyphen#1-#2\@nil{#2}%
  \def\secondofthree@hyphen#1#2#3\@nil{#2}%
  \def\submynthofm#1#2#3#4\@nil{\lastoftwo@hyphen#4\@nil}%
  \def\subsubmynthofm#1#2#3#4\@nil{\secondofthree@hyphen#4\@nil}%
}{}
\@ifpackageloaded{hyperref}{
  \newcommand{\subfirstofsix}[6]{\submynthofm#1\@nil}
  \newcommand{\subsubfirstofsix}[6]{\subsubmynthofm#1\@nil}
  \newcommand\subref[1]{\expandafter\real@setref\csname r@#1\endcsname\subfirstofsix{#1}}
  \newcommand\subsubref[1]{\expandafter\real@setref\csname r@#1\endcsname\subsubfirstofsix{#1}}
}{
  \newcommand{\subfirstofthree}[3]{\submynthofm#1\@nil}
  \newcommand{\subsubfirstofthree}[3]{\subsubmynthofm#1\@nil}
  \newcommand\subref[1]{\expandafter\@setref\csname r@#1\endcsname\subfirstofthree{#1}}
  \newcommand\subsubref[1]{\expandafter\@setref\csname r@#1\endcsname\subsubfirstofthree{#1}}
}
\@ifclassloaded{article}{

}{}
\@ifclassloaded{IEEEtran}{

}{}
\makeatother

\newcommand{\girder}[1]{
\titleformat{\subsection}{\normalfont\large\bfseries}{Girder \arabic{subsection}:}{0.5em}{}
\subsection{#1}
\titleformat{\subsection}{\normalfont\large\bfseries}{\thesubsection}{1em}{}
}

\definecolor{grey}{gray}{0.6}
\definecolor{darkred}{rgb}{0.44,0,0}
\definecolor{darkgreen}{rgb}{0,0.44,0}
\definecolor{darkblue}{rgb}{0,0,0.44}
\definecolor{enrique}{rgb}{0,0,0}

\definecolor{mygreen2}{HTML}{196B24}
\definecolor{myorange2}{HTML}{E97132}
\definecolor{myblue2}{HTML}{156082}

\usepackage{listings}
\definecolor{mygreen}{rgb}{0,0.6,0}
\definecolor{mygray}{rgb}{0.5,0.5,0.5}
\definecolor{mymauve}{rgb}{0.58,0,0.82}

\usepackage[normalem]{ulem}

\theoremstyle{definition}

{}
{}

{}
{}
{}

\pgfplotsset{compat=1.18}

\everymath{\displaystyle}
\usepackage{ifthen}
\usepackage{colortbl}
\usepackage{array}
\usepackage{color}
\usepackage{soul}
\usepackage{amssymb}

\newcolumntype{I}{!{\vrule width 1.5pt}}
\newlength\savedwidth

\newboolean{IsWide}
\setboolean{IsWide}{true}

\newcommand{\FlaPartition}[2]{
\ifthenelse{\boolean{IsWide}}{{\bf partition } \hspace{-1em} #1 \hspace{-1em} #2}
{{\bf partition } \+ \\ #1 \+ \\ #2 \- \-}
}

\newcommand{\FlaRepartition}[2]{
\ifthenelse{\boolean{IsWide}}{{\bf repartition } \hspace{-1em} #1 \hspace{-1em} #2}
{{\bf repartition } \+ \\ #1 \+ \\ #2 \- \-}
}

\newcommand{\rvdgFromTo}[2]{}{}

\newcounter{rivet}
\counterwithin*{rivet}{subsection}
\newcommand{\rivet}{\stepcounter{rivet}\item{\bf Rivet\kern3pt \arabic{subsection}.\therivet:\kern0.5em}}

\newcounter{metric}
\counterwithin*{metric}{subsection}
\newcommand{\metric}{\stepcounter{metric}{\bf Metric M\arabic{subsection}.\themetric}:~}
\WithSuffix\newcommand\metric*{\stepcounter{metric}{\bf Metric M\arabic{subsection}.\themetric}}
\newcommand{\milestone}{\stepcounter{metric}{\bf Milestone M\arabic{subsection}.\themetric}:~}
\WithSuffix\newcommand\milestone*{\stepcounter{metric}{\bf Milestone M\arabic{subsection}.\themetric}}

\renewcommand{\textcircled}[1]{\tikz[baseline=(char.base)]{
            \node[shape=circle,draw,inner sep=1pt] (char) {#1};}}

\newcommand{\NoShow}[1]{}

\newcounter{mycounter}

\newcommand{\commentout}[1]{}

\title{A Proposed Framework for Advanced (Multi)Linear Infrastructure in Engineering and Science (FAMLIES)%
\footnote{This work was supported in part by the National Science Foundation through the Cyberinfrastructure for Sustained Scientific Innovation (CSSI) program under NSF grants OAC-2513927, OAC-2513928, and OAC-2513929.

Any opinions, findings, and conclusions or recommendations expressed in this material are those of the author(s) and do not necessarily reflect the views of the National Science Foundation.}\\[0.2in]
\Large FAMLIES Working Note \#0}

\author{
Devin A. Matthews  \\
Department of Chemistry \\
Southern Methodist University \\[0.1in]
Tze Meng Low  \\
Electrical and Computer Engineering \\
Carnegie Mellon University \\[0.1in]
Margaret E. Myers \\
Devangi N. Parikh  \\
Robert A. van de Geijn  \\
Department of Computer Science \\
and  \\
Oden Institute for Computational Engineering and Sciences \\
The University of Texas at Austin
}
\begin{document}

\maketitle

\begin{abstract}
The Basic Linear Algebra Subprograms (BLAS), LAPACK, and their derivatives (PBLAS, ScaLAPACK, MAGMA, SLATE, etc.), which implement specific operations commonly encountered in dense linear algebra (DLA), have had an arguably unparalleled impact on scientific computing and, more recently,  machine learning and data science.
Part of their success initially was in the stringent enforcement of boundaries between layers (for example between single-node and multi-node levels or between BLAS and LAPACK-level functionality), via interfaces that are de facto standards.  
Over time, this has also become a weakness: the enforcement of these  boundaries is  now an impediment to reducing overhead due to data movement, be it within or between processing  units, and to the identification and exploitation of optimization opportunities such as loop fusion. 
Another challenge arises when adapting to new hardware architectures, such as graphics processing units (GPUs), while also quickly implementing new high-performance matrix and tensor algorithms that arise in scientific computing and data science. These challenges highlight the need for a more flexible approach in defining and implementing high-performance dense linear and tensor algorithms which can better adapt to changing applications, software, and hardware.

We leverage highly successful prior projects sponsored by multiple NSF grants and gifts from industry: the  BLAS-like Library Instantiation Software (BLIS) and the libflame efforts to lay the foundation for a new flexible framework by vertically integrating the dense linear and multi-linear (tensor) software stacks that are important to modern computing.   This vertical integration will enable high-performance computations from node-level to massively-parallel, and across both CPU and GPU architectures. The effort builds on decades of experience by the research team turning fundamental research on the systematic derivation of algorithms (the NSF-sponsored FLAME project) into practical software for this domain, targeting single and multi-core (BLIS, TBLIS, and libflame), GPU-accelerated (SuperMatrix), and massively parallel (PLAPACK, Elemental, and ROTE) compute environments. This project will implement key linear algebra and tensor operations which highlight the flexibility and effectiveness of the new framework, and set the stage for further work in broadening functionality and integration into diverse scientific and machine learning software.

\end{abstract}

\setcounter{page}{1}

\pagestyle{empty}

\section{Introduction}
\label{sec:introduction}

LAPACK~\cite{LAPACK3} and ScaLAPACK~\cite{ScaLAPACK}, first proposed in the early 1990's, have had a huge impact on scientific computing and, more recently, data analysis and machine learning.
Over time, derivatives like MAGMA~\cite{dghklty14} and PLASMA~\cite{PLASMA} addressed how to harness new advances in hardware such as GPUs and other accelerators.
Fundamental to all these efforts has been the strict adherence to layering, with the Basic Linear Algebra Subprograms (BLAS)~\cite{BLAS1,BLAS2,BLAS3}, standardized in the 1970's and 1980's, as the lowest layer that provides a level of readability (for those who are familiar with BLAS naming conventions) and performance portability across platforms.  Core to performance is the use of blocked or tiled algorithms that cast most computation in terms of matrix-matrix operations on sub-matrices (level-3 BLAS)~\cite{DDSV,LAPACK3}.

\subsection{Long-term vision}
\label{sec:vision}

The vision is to build on the vast experience from LAPACK and derivatives, our own research and development,  and other advances to create a flexible, modern framework for dense linear algebra (DLA) and multi-linear algebra (tensor) functionality for current and future compute platforms.

\subsection{Challenges}

Choices that were reasonable in the 1990's over time have become restricting given the heterogeneous nature of modern architectures and their deep memory hierarchies.  To name a few issues:
\begin{itemize}
\item
For a given DLA operation, a family of algorithms is needed so that the best can be chosen for a problem size, target hardware, and/or level of memory hierarchy.  The siloed approach to the coding of LAPACK already requires a huge code base.  Adding additional algorithms to this magnifies complexity, and increases the burden of individually optimizing each algorithm.
\item
Multiple levels of blocking are now required for near-optimal performance, requiring nested calls to the operation where at each level the best algorithmic variant is employed.  LAPACK and other libraries typically hard-code the number of levels and the algorithm used at each level.
\item 
Memory movement is the limiting factor for performance, leading to theoretical and practical advances regarding communication-avoiding algorithms~\cite{CommAvoiding0,CommAvoiding1,CommAvoiding2,CommAvoiding3,smith2019tightiolowerbound}.  The strict adherence to layering, with rigid interfaces like the BLAS, stands in the way of the fusing of operations so that memory movement can be reduced.
\item 
With the advent of new precisions like bfloat16 (bf16) and the exploitation of mixed precision to reduce computational cost, the space of operations and algorithms that need to be supported  leads to unmanageable complexity in libraries if LAPACK-like coding conventions are enforced~\cite{abdelfattah2020surveynumericalmethodsutilizing,10.1145/3402225}.
\item 
An important modern use of DLA libraries is in the context of tensor (multi-linear) computations.  Often, approaches leverage LAPACK and its derivatives by explicit conversion between tensors and matrices. Alternative approaches, for example based on fusion of data reorganization with  matrix-matrix multiplication~\cite{tblis_sisc}, can improve performance and more tightly integrate tensor structure. Matrix-centric algorithms and frameworks also hinder efficient higher-dimensional data distribution and complex operations such as tensor factorization.
\end{itemize}

\subsection{The solution: A vertically integrated framework for  (multi)linear algebra}

We expect a full realization of the vision to take a decade or more.  This project will lay the foundation for a new framework, FAMLIES, that overcomes  challenges through vertical integration, the flexible control of algorithms and communication, and a consistent programming API across levels. This framework will be designed for a broad, representative, and usable set of functionality that can be used instead of, or side by side with, current LAPACK-based products.

\subsection{Project Motivation}

Dense linear algebra and multi-linear algebra (tensor computations) are widely used in many scientific and machine learning workloads. Success in this effort will allow the quick instantiation of the appropriate algorithm for the different scientific and machine learning domains on different architectures and platforms. In particular, we are driven by the science in the fields of computational quantum chemistry and  machine learning. 

In quantum chemistry, DLA and tensor operations form the mathematical foundation of theories of electronic structure.  They are major computational bottlenecks. Quantum chemistry calculations are a major user of computing time on NSF and other national computing resources.  Accelerating both the pace of developing and implementing new electronic structure theories as well as  the speed at which calculations  run are major drivers of our approach and project goals.

Tensor operations are also core computational engines within many machine learning models.  Furthermore, many other machine learning computations exhibit similar, if not identical, data access patterns to those found in DLA and traditional scientific applications\cite{SMaLL,ICML}. The ability to quickly develop fast implementations of new algorithms will facilitate the exploration of new models for ML/AI. Different implementations of the same ML algorithms with different hardware requirements, and the ability to port them across different computing devices, will also speed up the deployment of these models on platforms ranging from data-center GPUs to IoT devices.

\begin{figure}[tb!]

\begin{center}
\begin{sideways}
\includegraphics[width= 0.9\textheight]{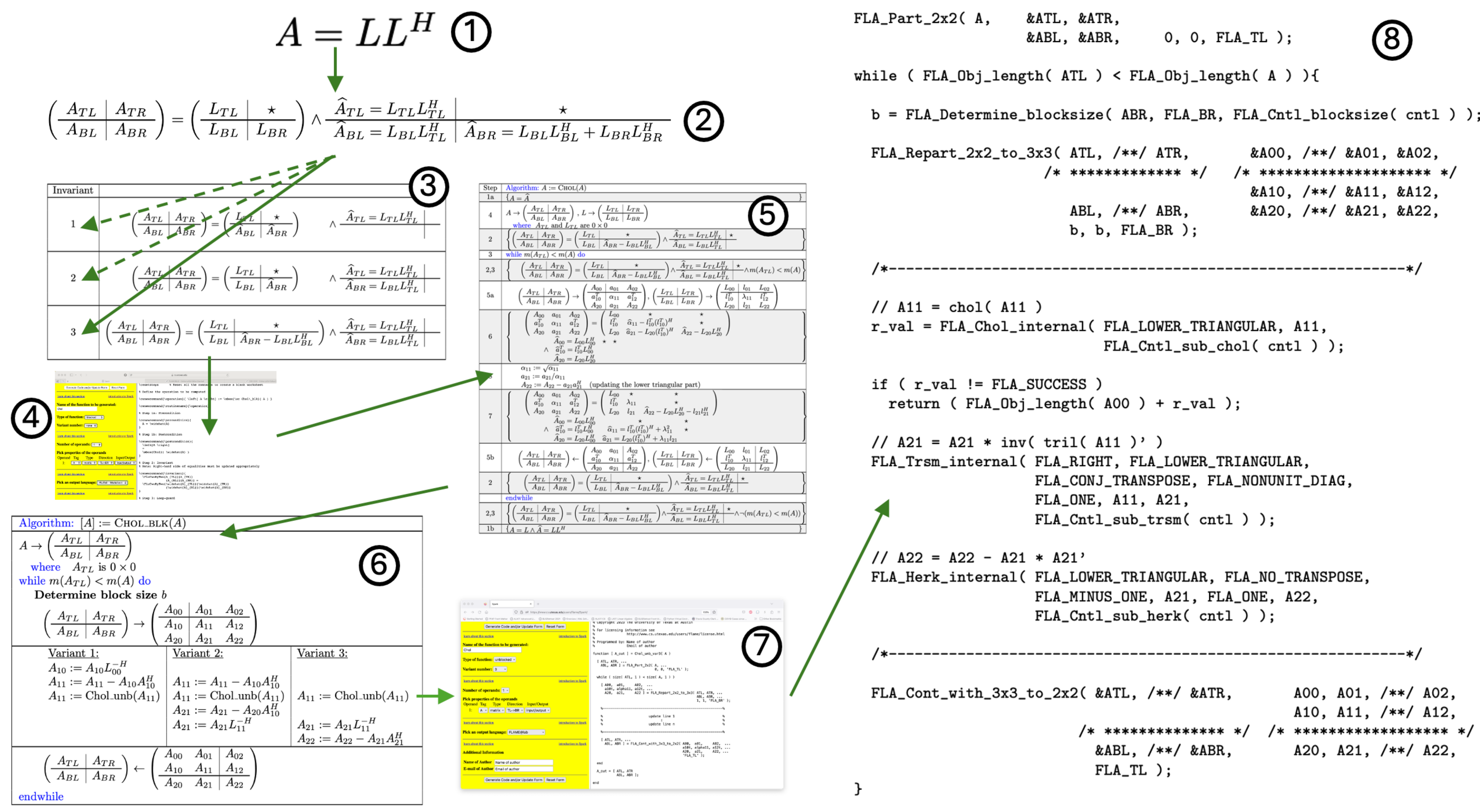}
\end{sideways}
\end{center}
\caption{The FLAME methodology workflow.}
\begin{comment}
\textcircled{{\footnotesize $1$}} 
    Start with the operation to be computed and derive the recursive definition (PME) for overwriting the matrix with its Cholesky factor; 
\textcircled{{\footnotesize $2$}}
From the PME, {\em a priori} derive feasible loop invariants;
\textcircled{{\footnotesize $3$}}
Pick a loop invariant and use the Spark webpage to generate a \LaTeX\ template for the worksheet;
\textcircled{{\footnotesize $4$}}
Use the worksheet to derive algorithmic variants;
\textcircled{{\footnotesize $5$}}
These result in multiple (in this case blocked) algorithmic variants;
\textcircled{{\footnotesize $6$}}
For each algorithmic variant, use the Spark webpage to generate a code skeleton using the FLAME API for the desired language; 
\textcircled{{\footnotesize $7$}}
Fill out the codes skeleton to reflect the updates in the derived algorithm.
\end{comment}
\label{fig:workflow}
\end{figure}

\subsection{Building on important  advances}
\label{sec:building}

Since the inception of LAPACK, sustained innovation by the PIs and their collaborators, as well as in the wider community, has led to the development of a number of critical components which motivate and facilitate the proposed work:

    \paragraph 
    {\bf Abstraction.}
    A key advance that enables rapid discovery of algorithms was the presentation of algorithms without explicit indexing, what we now call the {\em FLAME notation}~\cite{FLAME_WoCo,FLAME,inverse-siam}.  This is illustrated by \textcircled{6} in Fig.~\ref{fig:workflow} for the three blocked algorithmic variants for Cholesky factorizing.
    
    \paragraph 
    {\bf Deriving families of algorithms.}
    Embracing the FLAME notation has enabled the application of formal derivation techniques to this domain~\cite{FLAME_WoCo,FLAME_TR,Recipe,TSoPMC}.
   Using the Cholesky factorization in Fig.~\ref{fig:workflow}, one starts with \textcircled{1} the definition of the operation from which a recursive definition of the operation, the \textcircled{2} Partitioned Matrix Expression (PME), is derived.  From this, \textcircled{3} all loop invariants (describing variable states before and after each iteration) can be deduced.  \textcircled{4} A menu generates a \textcircled{5} worksheet outline, which is used to derive (hand in hand  with their proofs of correctness) \textcircled{6} algorithmic variants, summarized using the FLAME notation.  
    Whole families of algorithms for a broad range of DLA operations (within and beyond LAPACK) have been systematically derived~\cite{FLAME,TSoPMC,Recipe,Bientinesi:2008:FAR:1377603.1377606,10.1145/3544585.3544597}.
    
    \paragraph 
    {\bf Correctness in the presence of round-off error.}
    The FLAME methodology derives algorithms that are correct in  infinite precision.   In finite precision, correctness is captured by the backward or forward error analysis of an algorithm.  It has been shown that the FLAME methodology can derive such  error analyses~\cite{Bientinesi:2011:GMS:2078718.2078728}.  
    
    \paragraph 
    {\bf Representing algorithms in code.}
    FLAME notation can represent a broad cross section of DLA algorithms for functionality included  in (and beyond) LAPACK~\cite{FLAWN17,LTLt}, both known or newly derived via the FLAME methodology.  
    By adopting APIs that mirror the FLAME notation, correct algorithms can be translated to correct code \textcircled{7} by an automated system.  
    In Fig.~\ref{fig:workflow}, \textcircled{8} illustrates the FLAMEC API used by our  libflame DLA library~\cite{libflame_github,libflamebook,CiSE09}, which was funded  in part by a  NSF SI2 SSI  grant~\cite{SI2}.  
    Similar APIs were used by us for coding the distributed-memory DLA libraries PLAPACK~\cite{PLAPACK_SC97,PLAPACK} and Elemental~\cite{Elemental:TOMS}, and the distributed memory dense tensor contraction library  ROTE~\cite{ROTE}.  Tiled  algorithms~\cite{PLASMA,SuperMatrix:SPAA07,SuperMatrix:PPoPP09} (which we call  algorithms-by-blocks) that create Directed Acyclic Graphs (DAGs) of operations with blocks to be scheduled to multi-core and/or (GPU) accelerators are also coded this way in libflame~\cite{CiSE09,SuperMatrix:SPAA07,SuperMatrix:PPoPP09}.
    
    \paragraph
    {\bf BLIS and TBLIS: Building flexible frameworks for BLAS and beyond.}
    Our BLIS\break{}project\cite{BLIS1,BLIS2,BLIS3,BLIS4,BLIS5,BLIS6,SIAM-News-BLIS1,SIAM-News-BLIS2,blisweb} is an award-winning~\cite{SIAMSIGBestPaper,WilkinsonPrize}, widely-used open-source implementation of BLAS on CPUs, and a toolbox/framework for the rapid instantiation of BLAS\emph{-like} functionality. 
It was funded by two NSF CSSI grants~\cite{CSSI1,CSSI2} and gifts from industry. 
Articles in SIAM News provide details~\cite{SIAM-News-BLIS1,SIAM-News-BLIS2}.

\begin{figure}[tb!]
\begin{center}
\vspace{-1em}
\includegraphics[width=0.7\textwidth]{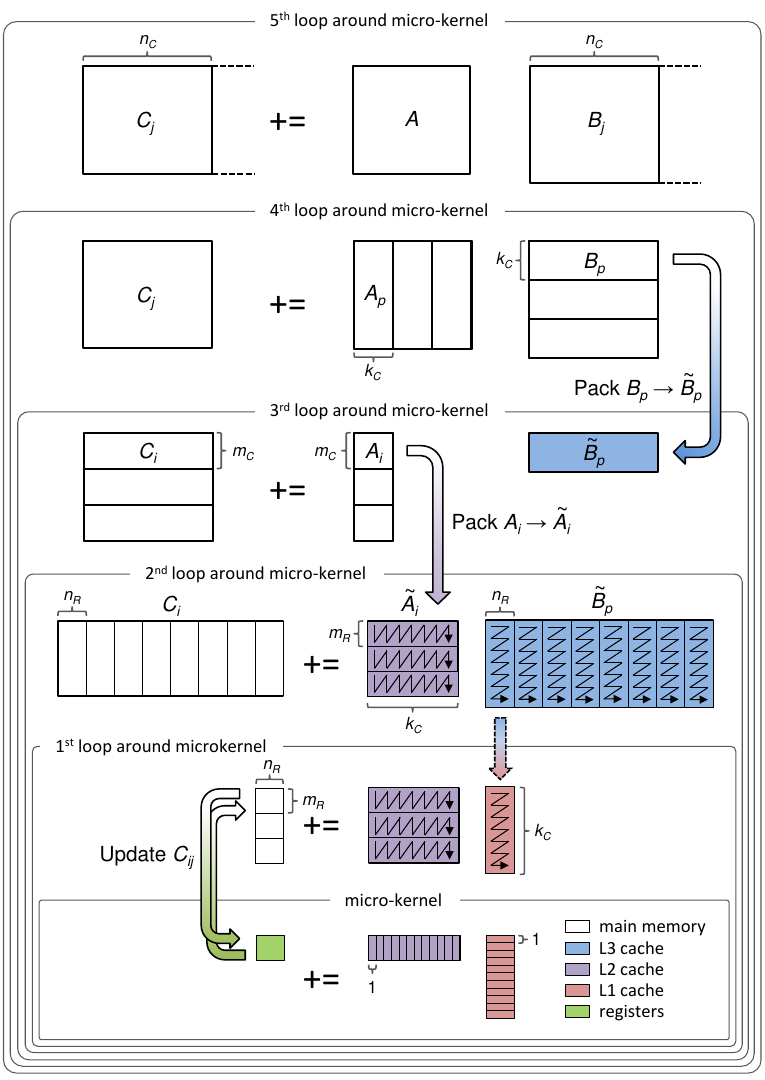}
\end{center}
\vspace{-3mm}
\caption{
The BLIS refactoring of the GotoBLAS algorithm as five loops around the micro-kernel.
% This diagram was modified from a similar image first published in~\cite{BLIS5} and is used with permission.
}
\label{fig:BLIS}
\end{figure}

With the 1.0 release, the portability of BLIS was extended to a wide variety of architectures.\footnote{x86-64 (Intel and AMD), ARM (arm32 and aarch64, esp. Ampere Altra), IBM POWER, RISC-V (esp. SiFive x280) and other architecture families.} BLIS's multi-threading capabilities were expanded to support diverse end-user applications and  scalability to hundreds of cores. It incorporates proposed changes~\cite{demmel2022proposed} to the BLAS that consistently handle exceptions like the propagation of {\tt NaN} and {\tt Inf}. 
It is included in AMD's Optimized CPU Libraries (AOCL) and the NVIDIA Performance Libraries (NVPL), and is packaged in Linux distributions.

Important to this proposal is that BLIS 2.0 allows easy extension of the BLAS to new operations with high performance by remixing BLIS's ``building blocks," illustrated in Fig.~\ref{fig:BLIS}, with user-specified components.  This has also been used to re-implement TBLIS, our high-performance tensor contraction library~\cite{tblis_sisc,TBLISweb}, to  be  leveraged by  the proposed work.

\paragraph{\bf  Nesting of algorithms and blockings.}
Improving performance requires careful composition of algorithms for the operations that together implement a given function.  
    
    For Cholesky factorization, at the top level, there is a choice of three blocked algorithms to be made, as well as the block size to be used at that level.  Each of these involves calls to level-3 BLAS as well as a recursive call to a Cholesky  factorization.  Hierarchically, choices of algorithm and block size are combined in multiple layers.  Together, this defines an enormous implementation space, especially if one also includes how to parallelize and how to redistribute data between memory layers, asymmetric compute resources, and nodes of a distributed-memory architecture.

    An innovation unique to libflame and BLIS is the {\em control tree}~\cite{libflamebook,blisweb}.  This is a hierarchical instruction that encodes the choices of algorithms and block sizes to be used in the implementation of a given function.  It manages the complexity of supporting the huge algorithmic space.
    
    \paragraph {\bf Controlling parallelism.}
    As modern heterogeneous and distributed architectures incorporate nested levels of parallelism, controlling that parallelism within an application and the libraries upon which it builds becomes paramount. 
    BLIS now supports {\em thread communicators} that provide such control within that layer of the software stack. This is a flexible abstraction that covers both thread-based and task-based parallelism with an MPI-like design.~\cite{BLIS3} %\textcolor{red}{TODO: cite BLIS threading paper}

      \paragraph 
     {\bf Reducing communication overhead.}
    Overhead due to data movement between memory layers and/or processing units has become more pronounced as the gap between the bandwidth to memory and the rate of computation has increased~\cite{CommAvoiding0,CommAvoiding1,CommAvoiding2,CommAvoiding3,smith2019tightiolowerbound}.  One solution is to embrace  communication-avoiding algorithms that achieve near-optimal amortization of computation over communication~\cite{CommAvoiding0,CommAvoiding1,CommAvoiding2,CommAvoiding3,smith2019tightiolowerbound}.  The  second is to fuse operations so as to reduce repeated data movement~\cite{SIAM-News-BLIS2,ltlt_perf}.

    \paragraph 
    {\bf Modeling performance.}  BLIS performance can be accurately and analytically  modeled~\cite{BLIS4}.  This allows blocking parameters to be calculated and supports choosing the best strategy from the space of algorithms supported by the proposed approach.~\cite{Strassen:SC16,FMM:IPDPS17} %\textcolor{red}{TODO: add Strassen/FMM paper citations}

    \begin{figure}[ptb]
{
\begin{lstlisting}[basicstyle=\ttfamily\footnotesize]
template <typename Type, int Variant, bool Blocked>
dim_t cholesky_l_impl(const marray_view<Type, 2>& A,
                      const control_tree& control)
{
    assert(A.length(0) == A.length(1));
 
    auto [T, B] = partition_columns(A, FORWARD);
 
    while (B)
    {
        auto [R0, R1, R2] = repartition<Blocked ? DYNAMIC : 1>
            (T, B, control.block_size(), FORWARD);
 
        if constexpr (Variant == 1)
        {
            // A10 = A10 * inv( tril( A00 )' )
            // RIGHT, LOWER, NONUNIT_DIAG encoded in control tree
            control.call<TRSM>(1, A[R0][R0], A[R1][R0].H());
 
            // A11 = A11 - A10 * A10'
            // LOWER encoded in control tree
            control.call<HERK>(-1, A[R1][R0], 1, A[R1][R1]);
 
            // A11 = chol( A11 )
            auto r_val = control.call<CHOL>(A[R1][R1]);
            if (r_val != SUCCESS) return T.size() + r_val;
        }
        else if constexpr (Variant == 2)
        {
            // ...
        }
        else /* if constexpr (Variant == 3) */
        {
            // ...
        }
 
        std::tie(T, B) = continue_with(R0, R1, R2, FORWARD);
    }
 
    return SUCCESS;
}
\end{lstlisting}
}
\caption{Prototype C++ implementation of the Cholesky factorization that illustrates design features contributing to vertical integration.}
\label{fig:cpp_chol}
\end{figure}

    %\paragraph
    %{\bf Randomized algorithms.}
%{\color{red} How does this fit in?  I am thinking  to  pull this out and somewhere  say this is orthogonal to what we are doing.   The framework should be such that randomized algorithms can, in principle, be implemented with it.}

    \paragraph 
    {\bf Supporting new and mixed precisions.}
    A key feature of modern architectures, and modern scientific and ML applications, is the introduction of new precisions.  A  major advance in BLIS has been the support of mixed-precision and mixed-domain computations across the level-3 BLAS, and continued work on introducing new precisions such as {\tt f16} and {\tt bf16}.~\cite{10.1145/3402225} %\textcolor{red}{TODO: cite BLIS MP/MD paper}
    
    \paragraph
    {\bf Exploiting modern C++ language features.}
    Using modern features of\break{}C++17 and later, we have developed expressive yet highly efficient facilities for working with vectors, matrices, and tensors in our TBLIS\cite{tblis_sisc,TBLISweb} and MArray\cite{marray} libraries. These interfaces support the rapid, efficient, and user-friendly implementation of complex DLA operations.\cite{ltlt_perf}
    
    \paragraph 
    {\bf Lowering barriers.}  Over the last decade, our team has developed four Massive Open Online Courses (MOOCs)~\cite{ulaff} and related materials~\cite{LAFF,LAFF-On-Correctness,LAFF-On-HP,ALAFF-book} that are offered on the edX platform~\cite{edX} (for free to auditors) and as in-class and online courses at UT Austin.  These courses link undergraduate and graduate level linear algebra to their high-performance implementation using the FLAME abstractions and methodologies, thus lowering barriers to entry into the field.

    \paragraph 
    {\bf Engaging a community.}
    The BLIS project has a very vibrant community of users and contributors from academia and industry.  This encompasses monthly advisory meetings, mailing lists, yearly workshops with stakeholders (BLIS Retreats~\cite{blis_retreat}), GitHub project,\cite{blisweb} and an active Discord server~\cite{BLISDiscord}.  

\vspace{0.1in}
Our decades of experiences tell us that these and other advances will allow us to achieve the stated goals while managing complexity.

\section{Approach}
\label{sec:approach}

\begin{comment}
    It may be worthwhile to compare and contrast this code with that of LAPACK and its various derivatives%
\footnote{
\begin{tabular}[t]{@{}l}
LAPACK: \url{https://netlib.org/lapack/explore-3.1.1-html/dpotrf.f.html},  \\
ScaLAPACK: \url{https://www.netlib.org/scalapack/explore-html/d5/d9e/pdpotrf_8f_source.html}.
\end{tabular}
}, 
and FLAME derivatives
\footnote{
\begin{tabular}[t]{@{}l}
libflame:, 
\tiny
\url{https://github.com/flame/libflame/blob/master/src/lapack/dec/chol/l/flamec/FLA_Chol_l_blk_var3.c}.
\end{tabular}}.
\end{comment}

We now detail some of the key novel ideas that underlie our proposed framework. These extend the long history of innovation summarized in Section~\ref{sec:building} and illustrate the feasibility and potential benefits of vertical integration of the  software stack. Some illustrations of the performance benefits enabled by our approach are reproduced in Fig.~\ref{fig:ltlt}.

\subsection{Implementing a space of algorithms for each operation.}

We use the prototype C++ implementation of Cholesky factorization in Figure~\ref{fig:cpp_chol} to illustrate how vertical integration of the dense linear algebra software stack can be achieved.  The full details of what code will look like will be determined as the project progresses.  

The following observations point to how a framework yields a enormous reduction in lines of code while simultaneously encoding a large space of algorithms with  the code  in Figure~\ref{fig:cpp_chol}:

\paragraph{\bf Variations on functionality.} It implements both $ A \rightarrow L L^T$ and $ A \rightarrow U^T U $ by implicitly transposing (switching the strides between row and column elements).

    \paragraph{\bf Data types.}
    The code in itself describes the mathematical computations that need to be performed. Hence, it supports
    %In principle, the code supports
    all precisions (single, double, half,  ...) and domains (real, complex). Mixtures of data types can be achieved by introducing more type parameters.
    
\begin{figure}[tb!]
\begin{center}
%\begin{tabular}{p{0.25\textwidth} p{0.33\textwidth}}
\includegraphics[width=0.4\textwidth]{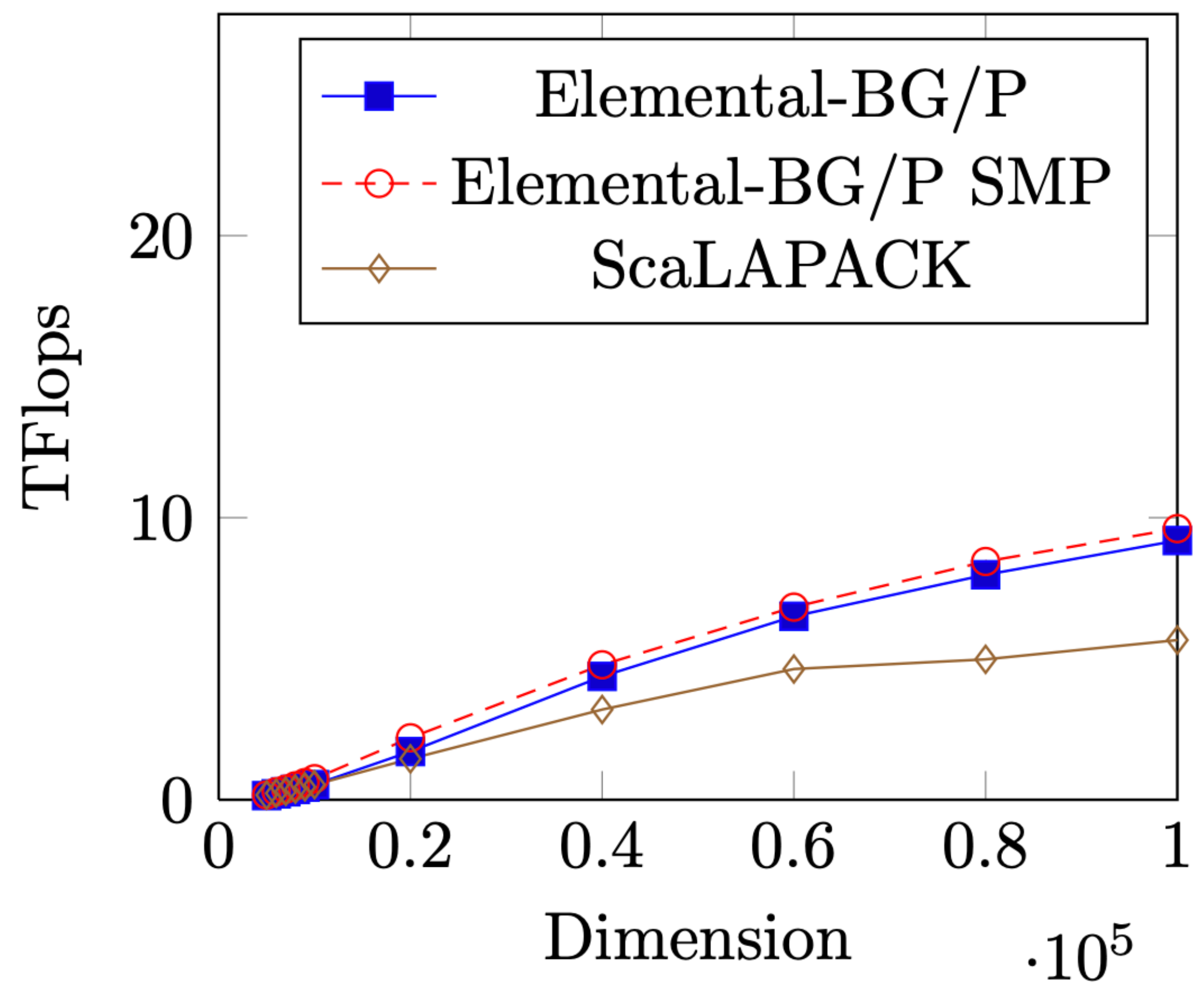} \quad
\includegraphics[width=0.55\textwidth]{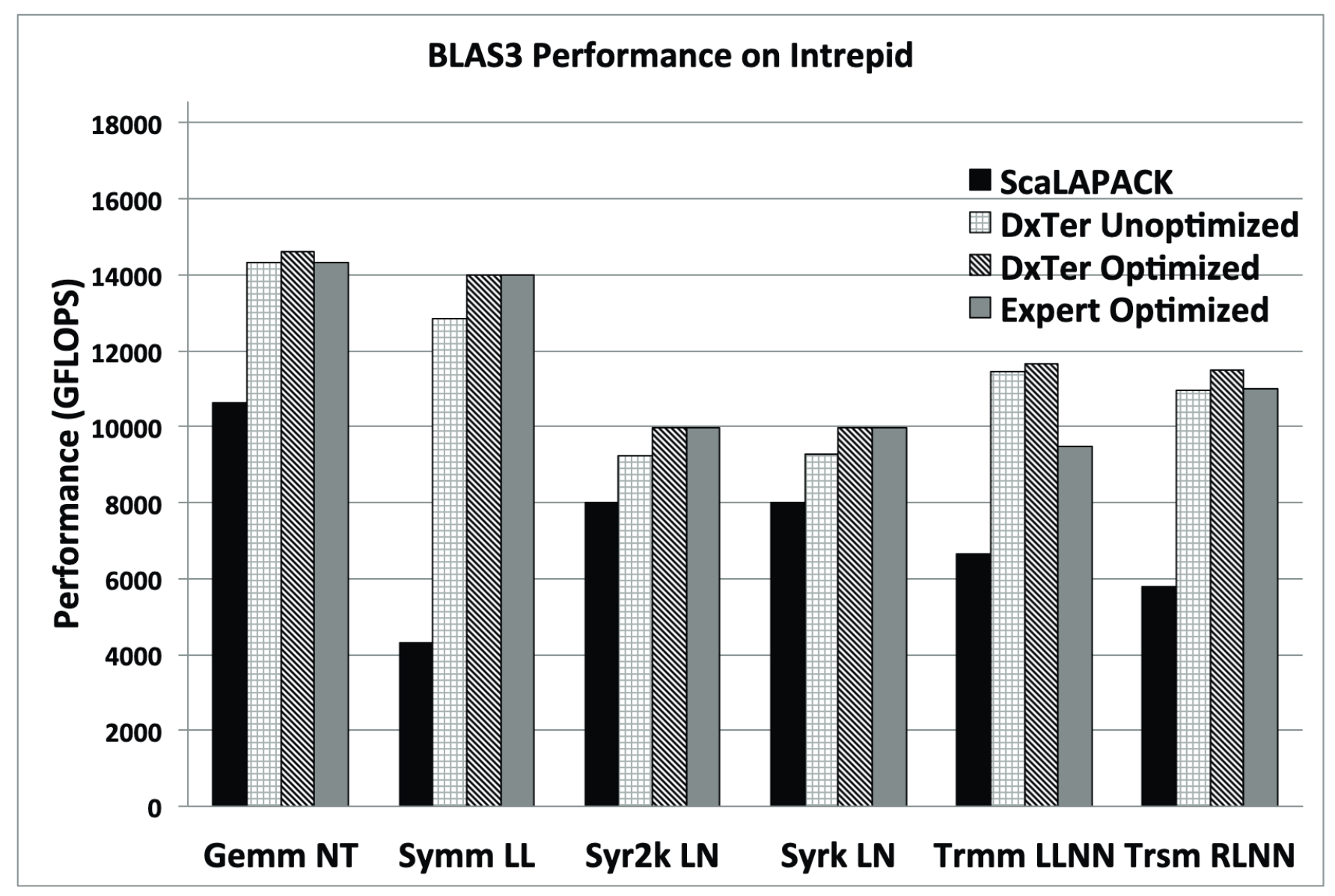} \\[0.2in]
\includegraphics[width=0.8\textwidth,height=7cm]{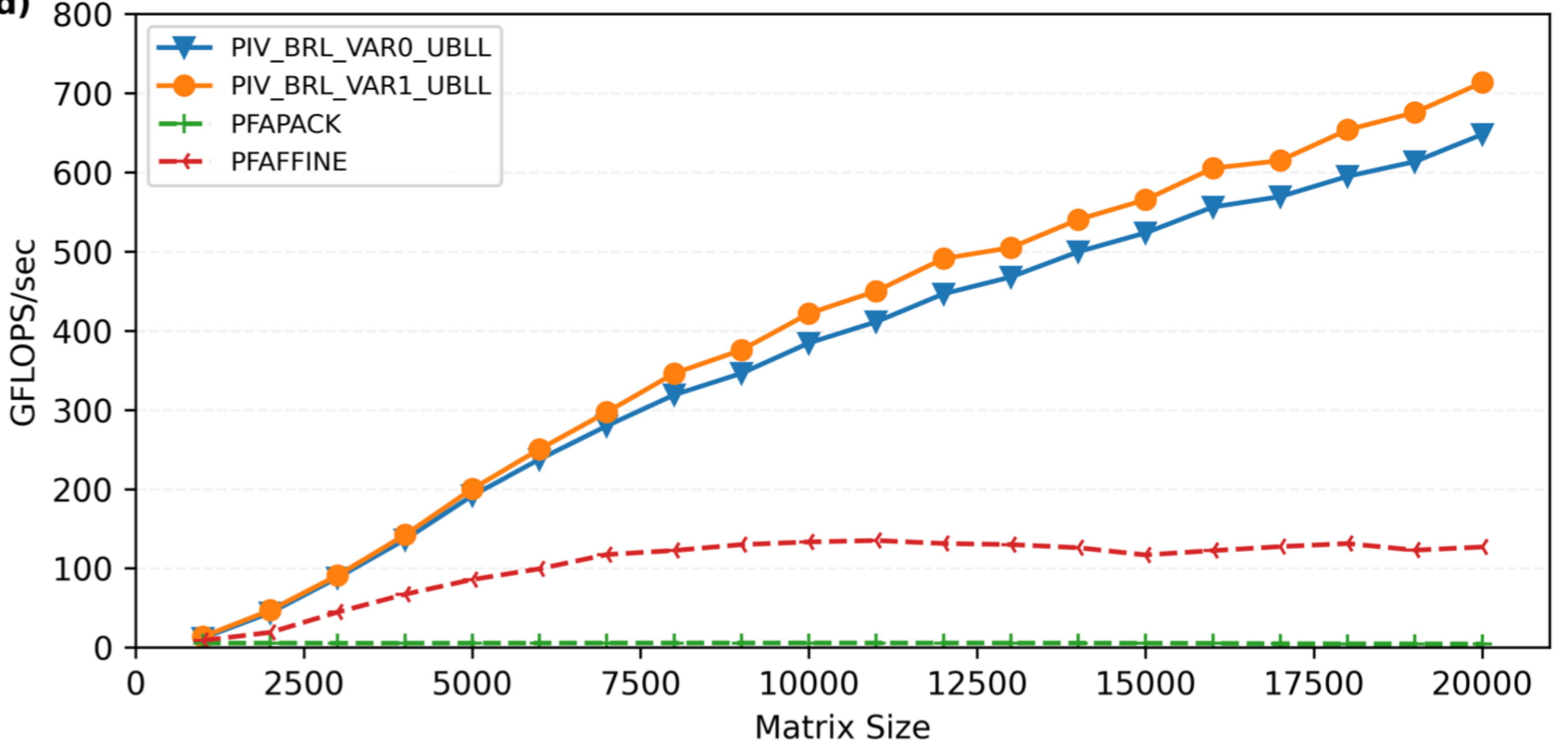}
%\end{tabular}
\end{center}
\caption{
Top-left: Comparison of performance of Elemental vs. ScaLAPACK on 8192 nodes of Argonne's IBM BlueGene/P system.\cite{Elemental:TOMS}. Top-right: Comparison of ScaLAPACK, hand-optimized Elemental, and a search over families of algorithms defined by composition (DxTer).\cite{iWAPT13}
Bottom: Performance of two algorithmic variants (top two curves) vs. best prior work (bottom two curves), implemented with the kind of techniques described in this proposal, for computing $ P X P^T = L T L^T $ on an AMD EPYC 7763 processor (64 cores).\cite{ltlt_perf}.
}
\label{fig:ltlt}
\end{figure}

    \paragraph{\bf Algorithmic variants.}
    Like the libflame code in \textcircled{8} of Figure~\ref{fig:workflow}, the code captures the algorithms  in
    \textcircled{6} of that figure, all in one implementation. The generation and exploration of families of algorithmic variants is key to discovering novel algorithms and for tailoring algorithms to specific hardware or problems. An example of how the automation of this process can exceed the performance of hand-optimized code is given in Fig.~\ref{fig:ltlt}[center].
    
    \paragraph{\bf Flexible abstractions.}
    It uses the {\em range} abstraction of MArray\cite{marray} to capture parts of the matrix (with {\tt R0}, {\tt R1}, and {\tt R2}). This is important for a number of reasons: (1) it removes all overhead associated with the abstractions used by libflame in \textcircled{8} of Figure~\ref{fig:workflow}; (2) it encodes both the blocked and unblocked implementations%
        \footnote{Examining the resulting compiled code shows no noticeble overhead relative to the unblocked algorithm that uses explicit indexing and calls to level-2 BLAS that is in, for example, LAPACK.}; (3) in algorithms involving multiple matrices and/or vectors, it links the partitioning of dimensions%
        \footnote{For example, consider $ C = A B $.  The partitioning of rows of $ C $ and $ A $, columns of $ C $ and $ B $, and the ``inner dimension'' of $ A $ and $ B $, are typically conformal, which can be indicated by using the same partitioned ranges for such pairs.}; (4) this range abstraction allows FLAME-like APIs to be used for tensor algorithms.  
    
    \paragraph{\bf Layering algorithms.}
    The code implements an entire space of algorithms for computing the Cholesky factorization via the control tree {\tt control} that is passed in.  In the proposed framework, this control tree will span all levels of the algorithms and tie together heterogeneous architectures and levels of parallelism.

\subsection{Beyond LAPACK functionality}

As an example of functionality beyond that covered by traditional LAPACK, consider skew-symmetric matrix factorizations.
Skew-symmetric matrices are encountered in diverse fields such as machine learning\cite{liu_deep_2017}, physics\cite{thomas_exact_2009}, and quantum chemistry/materials science\cite{bajdich_electronic_2010,xu_optimized_2022}. A central quantity is the Pfaffian of a skew-symmetric matrix $ X $, which can be computed by factoring $ P X P^T = L T L^T $, where $ P $ is a permutation matrix, $ L $ is unit lower triangular, 
%that accumulates Gauss transforms
and $ T $ is tridiagonal skew-symmetric.

Recent work of ours~\cite{LTLt,ltlt_perf} yielded new algorithms and implementations for this operation  that attain higher performance than the  best prior work~\cite{Wimmer2012}. This study tells us that:
\begin{itemize}
\item 
The application of the FLAME workflow in Figure~\ref{fig:workflow} to this new operation yields both known and new unblocked and blocked algorithms.
\item 
Performance is improved by the implementation with BLIS of new ``sandwiched'' matrix multiplication operations like $ C  := C - A T  A^T $, where $ C $ and $ T $ are skew-symmetric and tridiagonal skew-symmetric matrices, respectively%
\footnote{If cast in term of traditional BLAS, this would involve the computation of $ B = T A^T $ followed by the update $  C := C - A B$, updating only the lower triangular part of $ C $.  This second operation is known as GEMMT, the only extension of the traditional BLAS to have caught on since their original specification.
By integrating the formation of $ T A^T $ into the packing of $ B $ in Figure~\ref{fig:BLIS}, memory movement is reduced and workspace is avoided, yielding better performance.
}.
\item 
The Cholesky factorization in Figure~\ref{fig:workflow} utilize a ``$ 2 \times 2 $ to $ 3 \times 3 $'' repartitioning. Operations involving tridiagonal matrices require more parts to be exposed, necessitating a ``$ 3 \times 3 $ to $ 4 \times 4 $'' or ``$ 3 \times 3 $ to $ 5 \times 5 $'' (for unblocked and blocked algorithms, respectively) repartitioning to be employed.
This is elegantly supported by the proposed use of ranges in C++~\cite{ltlt_perf}.
\item A flexible and expressive API (similar to Fig.~\ref{fig:cpp_chol}) provides high performance (Fig.~\ref{fig:ltlt}[right]) and the ability to not only support traditional functionality, but to also provide components with which to assemble new functionalities.
\end{itemize}

\subsection{Supporting tensor computations}

Tensor computations extend DLA into the realm of multi-linear algebra. Essentially, tensors are collections of structured, multidimensional data, with the common case of dense tensors directly representable as multidimensional arrays. 
Tensors are critical in several scientific fields such as quantum chemistry and machine learning, where they represent quantities such as electronic and nuclear wavefunctions, batches of images, multi-head attention embeddings, and other multi-dimensional data. 
Based on our previous work on tensors and tensor algorithms,~\cite{tblis_sisc,10.1145/3301319,doi:10.1137/17M1135578,STTM,ROTE,6569864}
%\textcolor{red}{CITE: TBLIS, spin-summation paper with Paul, tensor Strassen, CTF with Edgar, ROTE/Martin symmetric tensor stuff, others?} 
we have developed several key techniques which support the proposed work:
\begin{itemize}
    \item The layering of algorithms inherent to the exploitation of levels of cache and/or distributed processor grids can be naturally extended to tensors by including additional layers for higher tensor dimensions. While it is possible to reduce tensors to matrices and then use matrix algorithms (e.g. the ``loop over {\sc gemm}'' or LoG approach),\cite{di_napoli_towards_2014,li_input-adaptive_2015} the use of recursive, layered algorithms in higher dimensions allows for a more diverse family of algorithms and better opportunities for optimization, e.g. when LoG would lead to non-contiguous data access.
    \item In TBLIS\cite{tblis_sisc}, we employed a ``block-scatter'' tensor-to-matrix mapping. This concept allows for matrix algorithms to be used directly on tensors, with an additional level of indirection for access to data based on indexing vectors. This approach is extensible to functionality across the levels considered here, as well as to alternative, operation-specific tensor-to-matrix mappings.
    \item Techniques for distributing matrices across two-dimensional processor grids extend naturally to tensors when the processors are viewed as a grid of the same dimensionality as the data. However, a more flexible and extensible technique is based on the concept of index filters~\cite{Elemental:TOMS,ROTE}.
    %\textcolor{red}{MArtin's thesis?} 
    This allows data of varying dimensionality to be distributed over common processor grid(s) and generalizes tensor redistributions using well-defined communication patterns as in Elemental~\cite{Elemental:TOMS} and ROTE~\cite{ROTE}.
    \item Tiling of tensors has long been used as a means to control the distribution and communication of tensor data and ensure sufficient work on-node, using a ``tensor-of-tensors'' approach where tiles are treated as discrete, persistent units\cite{tce,tiledarrays,exatensor,tamm}. A more flexible approach is to define tiles dynamically via blocking along one or more tensor dimensions. This approach enables customization of the data layout and communication patterns to the specific operation, reducing communication overheads both on-node and between nodes.
\end{itemize}
Additionally, coauthor Matthews has been actively involved in interdisciplinary efforts at tensor interface standardization, the most effective of which has been the Tensor Algebra Processing Primitives (TAPP) interface which arose out of a recent meeting organized by CECAM.\cite{cecam_tensor}

\subsection{Supporting the hardware stack}

PLAPACK~\cite{PLAPACK} was proposed in the late 1990's as a framework for implementing LAPACK-like functionality on distributed memory architectures, as an alternative to ScaLAPACK.
 It exposed that collective  communication patterns that are fundamental to DLA~\cite{InterCol:CCPE,SUMMA:SISC} could not be supported by ScaLAPACK due to design decisions underlying its use of the Basic Linear Algebra Communication Subprograms (BLACS)~\cite{DMCC6:BLACS,2dblacs,BLACS2} and parallel BLAS (PBLAS)~\cite{PBLAS}.
PLAPACK used object-based programming inspired by MPI~\cite{MPI1,MPI2} to overcome the complexity of managing indices at the local and global matrix level, a precursor to what became the FLAME APIs. Later, a modern instantiation of these ideas became the Elemental library,\cite{Elemental:TOMS} which significantly out-performs previous implementations (Fig.~\ref{fig:ltlt}[left]).
Key to both PLAPACK and Elemental was the inlining of data movement by describing what redistribution/reduction of data was required, where communication was hiding in calls that achieved those redistributions/reductions.

When accelerators like GPUs became popular, it was recognized that tiled algorithms~\cite{PLASMA,SuperMatrix:SPAA07,SuperMatrix:PPoPP09}  allowed a separation of concerns between algorithms that create a Directed Acyclic Graph (DAG) of operations with tiles of the operands and a runtime that manages the dispatching of data to resources for execution.  Importantly, in our libflame library, such algorithms are encoded with a combination of code that looks like that in Figure~\ref{fig:workflow} and clever use of the control tree~\cite{libflamebook}. 

Our experience is that
     data movement (rearranging, duplicating, and reducing) can  be elegantly added to code like that in Fig.~\ref{fig:cpp_chol}  to support the packing for data locality/redistribution/reduction for the efficient use of a single node,  NUMA, accelerator, and/or distributed memory architecture.  

\section{Proposed Work}
\label{sec:work}

Briefly, the goal of the work is to lay the foundation for a framework that vertically integrates the dense linear and multi-linear (tensor) software stack to support functionality spanning BLAS- and LAPACK-level (and beyond) dense linear algebra as well as multi-linear tensor computations, on scales from sequential or shared-memory multiprocessing to exascale distributed computation, and leveraging both CPU and GPU/accelerator resources. Upon completion, the project will have demonstrated that the framework, with additional contributions, can broadly support  functionality by providing families of implementations for a range operations on a range of  architectures.

Due to the cross-cutting nature of this vertical integration, we organize the project goals into ``girders'', which span both vertically (e.g. sequential to distributed parallel or BLAS to LAPACK) and horizontally (e.g. across DLA and tensor functionality) to form a strong yet flexible framework. 
Specific work items are represented as ``rivets'' which punctuate these activities and tie different girders together. Design goals for each girder are given.

\begin{comment}
This creates three dimensions of a space to be explored as part of this demonstration:
\begin{itemize}
    \item 
    {\bf Functionality:}  We will sample a range of functionality, including 
    \begin{itemize}
        \item 
        Linear solvers based on the ``Three Amigos'' (the LU with pivoting, Cholesky, and Householder QR factorizations).
        \item 
        Eigensolvers.
        \item 
        Singular Value Decomposition (SVD) related functionality.
        \item 
        Beyond LAPACK DLA functionality.
        $ L T L^T $ etc.
        \item
        Tensor contraction with general and  symmetric  dense tensors.
    \end{itemize}
    \item 
    {\bf Families of algorithms:}
    For each operation,  families of algorithms will be included.  This will encompass variants derived using the FLAME workflow as well as alternative formulations (e.g., algorithms  based on the QR algorithm and method of Multiple Relatively Robust Representations (MRRR) for the symmetric eigenvalue problem.
    \item 
    {\bf Architecure:}
\end{itemize}
\end{comment}

\girder{Consistent Application Programming Interface}\label{girder:API}

We target a flexible framework which breaks through traditional layering and separation of APIs by functionality and architecture. While such traditional APIs can still be defined (see Girder~\subref{girder:legacy}) to encompass fixed functionality, we focus on a portable ``algorithmic'' API to support the actual implementation of such functionality.

\paragraph{Design Goals:}
\begin{itemize}

\item{\bf Ease of use:} A consistent interface across layers and architectures will enable users to easily experiment with new algorithms while also readily exploiting optimized kernels and other primitives, scaling up from shared-memory to distributed parallelism, and leveraging GPU acceleration without artificial barriers. We will evaluate ease-of-use through our own implementation of important DLA and other operations (see Girder~\subref{girder:functionality}) along with external evaluation by community members. % % \metric community surveys will rate the usability of the API. \metric case studies with community volunteers will expose areas for improvement.

\begin{comment}
\item{\bf Code generation:} In order to better aid exploration of the space of possible algorithms (see Girder~\subref{girder:algorithm}), automated generation of algorithms using the FAMLIES framework is a top priority. While code generation approaches do not necessarily require a consistent and user-friendly interface as described here, these features aid in extensibility of the code generation framework and the transparency and maintainability of the generated code. {\color{red} [MAYBE JUST DELETE THIS...]}
\end{comment}

%{\color{magenta}{Are we talking about code generation on steriods? Most code generators only target a small subset of architectures/platforms. Do we intend to have a code generator to generate code from CPU to GPU to Distributed systems of heterogeneous architecture? 

%Is this similar to Saman/Jon/Federick's approach of separating computation from schedule?
%}}

\item{\bf Enabling optimizations:} While core numerical kernels are often written in assembly or compiler intrinsics, the bulk of the algorithms in FAMLIES will be expressed using the general algorithmic API. 
On GPUs, the algorithm API may often extend all the way to the generation of scalar code. 
Thus, efficiency of lower levels of the algorithm critically depends on optimizations such as loop fusion and inlining. 
Additionally, the API must be flexible enough to incorporate a wide array of kernels and operation types which arise due to loop fusion. 
% \metric compiler optimization of unblocked CPU and GPU algorithms approaches that of existing ``hand-written'' implementations, based on micro-benchmarks and disassembly analysis.

\item{\bf Code safety: } While information security in scientific applications is often overlooked, hardening applications and numerical libraries against attack vectors such as buffer overflow (more generally, out-of-bounds access), use-after-free, and other memory safety issues is critical for safe usage in critical infrastructure. Additionally, graceful error recovery is important to maintain uptime and prevent denial of service attacks. Modern C++ provides many mechanisms for enforcing memory safety. 
% \metric The API supports end-to-end bounds checking (algorithmic and data access bounds). \metric Preferred error-handling mechanisms and bounds on exceptional behavior will be determined in collaboration with community members.

\end{itemize}

\paragraph{Project Plans:}

\begin{itemize}
\rivet A consistent API which can be used at multiple levels (unblocked, blocked, distributed) and across architectures (CPU, GPU) will be defined. 
% \milestone a preliminary interface covering all architectures and execution levels, with refinement to follow. See Girder~\subref{girder:algorithm} for complementary milestones relating to the representation and manipulation of entire algorithms.
\rivet How best to generalize concepts arising across levels, while also enabling customization where applicable, will be explored. 
% \milestone 
The API will generalize across BLAS, LAPACK, ScaLAPACK, and multi-linear (tensor) layers. 
%See Girder~\subref{girder:communication} for complementary milestones relating to the representation of communication and data layouts and Girder~\subref{girder:functionality} for milestones related to specific functionality.
\rivet 
How compiler optimizations, bounds-checking, error handling, loop fusion, and other optimizations can be enabled and exploited across levels using an object-based C++ approach will be explored. 
%\milestone implementation of robust error handling based on metric M1.5. \milestone we will promote and verify compiler-level optimization of code (particular low-level unblocked code) using the internal API. See Girder~\subref{girder:algorithm} for complementary milestones relating to optimization via algorithm transformation.

\end{itemize}

\girder{Comprehensive Control of Communication}\label{girder:communication}

The ever-increasing gap between time and energy required for communication (data movement) compared to computation (arithmetic operations) has increasingly driven the design of high-\break{}performance numerical algorithms. Controlling the flow of data, and how data is packed and stored across all levels of the memory hierarchy is a critical feature of the proposed framework. Different architectures and scales of computation have traditionally exploited fairly distinct forms of data flow control, for example temporal access locality and prefetching on CPUs, explicit collective or point-to-point communication (message passing) in distributed computation, and use of local shared memory and registers on GPUs. In this proposal, we adopt an approach which leverages vertical integration of the software stack.

\paragraph{Design Goals:}
\begin{itemize}
\item{\bf Exposing commonalities of data movement:} We seek to expose commonalities within and between different scales and architectures, while leveraging existing technologies (e.g. GPU-aware MPI~\cite{MPI1,MPI2}
%\textcolor{red}{[CITE?]}) 
on the back-end). In particular, we will emphasize the close connection between data movement and choice of algorithm, extending ideas from Elemental,\cite{Elemental:TOMS}%\textcolor{red}{[CITE]} 
ROTE,\cite{ROTE} %\textcolor{red}{[CITE]} 
CUTLASS,\cite{cutlass}
%\textcolor{red}{[CITE]} 
CuTe,\cite{cutlass-cute}
%\textcolor{red}{[CITE]} 
BLIS thread communicators,~\cite{BLIS3}
%\textcolor{red}{[CITE]} 
and others. 

\item{\bf Providing flexibility in data layout:} To enable clean and efficient implementations (Girder~\subref{girder:legacy}) and to explore of a broad algorithmic space (Girder~\subref{girder:algorithm}), it is important to be able to deal with a diverse set of data layouts, particularly at the distributed level. % \metric support for cyclic, block cyclic, generalized Morton~\cite{FLASH:TR} and (non-uniform) blocked distributed layouts. \metric support for arbitrarily-strided local matrix and tensor layouts.

\item{\bf Enabling optimizations across layers:} Reducing data movement is a key tool for increasing performance, and vertical integration can enable additional data movement optimizations. For example, packing for BLAS-level operations can be hoisted into the LAPACK (or multi-linear algebra) layer, and even into distributed communication. Comprehensive control of data movement across layers enables such optimizations. 
% \metric the performance impact of individual optimizations or optimization strategies will be measured on targeted functionality (Girder~\subref{girder:functionality}) and in domain-specific applications as appropriate.
\end{itemize}

\paragraph{\bf Project Plans:}

\begin{itemize}
\rivet A consistent framework API which can be used to describe data layouts at  multiple levels (unblocked, blocked, NUMA, distributed) and across architectures (CPU, GPU) will be defined. 
% \milestone a preliminary interface covering all architectures and execution levels, with refinement to follow. %\metric t
The interface should allow for a natural expression techniques such as recursive blocking and 3D (2.5D) algorithms which approach communication lower bounds.\cite{CommAvoiding0,CommAvoiding1,CommAvoiding2,CommAvoiding3} 
%\textcolor{red}{[CITES (Demmel for recursive CHolesky I think)]} 
% See Girder~\subref{girder:API} for complementary milestones related to the framework API.
\rivet 
Back-end libraries and/or computational kernels for relevant communication patterns (distributed collective and point-to-point communication, pre-packing, tensor redistribution, etc.) will be integrated. % \milestone on-node CPU and GPU data redistribution. \milestone distributed collective and point-to-point communication. See Girder~\subref{girder:algorithm} for milestones related to runtime scheduling into which communication will be incorporated.
\end{itemize}

\girder{Algorithm Control and Exploration of the Algorithmic Space}\label{girder:algorithm}

%Control tree: what does it look like? (templates, code generation/DSL, runtime; doesn't have to be one thing), generalize to tensors/new functionality, space of kernels/operations (how flexible can this be?), constraints on design space (theoretical, empirical, practical), fine-tuning, FUSION!!!

The design of modern high performance frameworks is often built on individual code kernels that are optimized for various use cases and serve different functions~\cite{BLIS1, SMaLL, FFTW}.  A single implementation of an algorithm within such a framework is obtained by the composition of kernels. Different combinations of these kernels provide a search space of algorithms and implementations for the same numerical operation~\cite{libflamebook}. Each of these algorithmic implementations has its own performance signature due to inter-processor communication, local data movement, workspace requirement, and execution details~\cite{DxTStairs}.  Furthermore, using kernels that target different architectures facilitates portability~\cite{BLIS2,SMaLL,ME}.
In this girder, we focus on designing the framework for integrating the outputs of Girders 1 and 2.
%The vertically integrated framework that cuts across both 

\paragraph{Design Goals:}
\begin{itemize}
\item{\bf Enabling optimizations:} Given the large space of algorithms and implementations for a single operation, the proposed framework has to be flexible enough to yield optimal or near-optimal algorithms based on specific hardware and problem specifications. The framework needs to be able to represent diverse algorithms, provide high-performance implementations, expose optimization opportunities such as fusion, and finally enable selection of an appropriate algorithm for given circumstances.

\item{\bf Leveraging existing frameworks:} Much previous work has identified and implemented different code kernels that support data (re)packing and computation for various architectures~\cite{BLIS1,SMaLL,tblis_sisc}. The proposed framework should  facilitate the use of existing kernels in various routines in order to leverage previous contributions to frameworks such as BLIS. The framework should also enable emerging techniques such as randomized sketching.\cite{hqrrp}
%\textcolor{red}{CITE}

\item{\bf Exposing parallelism explicitly:} Different architectures require different levels of parallelism and this available parallelism needs to be  exploited in different forms. The framework should allow different run-time systems (e.g. IRIS~\cite{10605063}, SuperMatrix\cite{SuperMatrix:PPoPP09}), and different parallelism frameworks (e.g. OpenMP, MPI, CUDA/HIP) to be used.

\end{itemize}

\paragraph{\bf Project Plans:}

\begin{itemize}
\rivet A consistent framework API for introducing and composing kernels and other building blocks into algorithms of interest to the (multi-)linear algebra communities will be defined. % \milestone incorporation of BLIS kernels. \milestone incorporation of TBLIS kernels. \milestone incorporation of other kernels of interest (such as fused kernels, randomized algorithms, efficient sequences of Givens rotations,\cite{fieldGivens,steel2022novel} 
%\textcolor{red}{CITE FIELD AND/OR THIJS?} and activation functions).
%Milestone M2.5: a preliminary interface covering all archi-
%tectures and execution levels, with refinement to follow. %Metric M2.6: the interface
%should allow for a natural expression techniques such as recursive blocking and 3D
%(2.5D) algorithms which obtain (approach) communication lower bounds.[CITES
%(Demmel for recursive CHolesky I think)] See Girder 1 for complementary mile-
%stones related to the framework API
\rivet The BLIS and libflame control trees across all levels and architectures will be combined and extended. Several complementary design strategies such as runtime structures, code generation, (embedded) domain-specific languages (DSLs), and templates (meta-programming) and combine them as appropriate for architecture and performance constraints will be explored. % \milestone representation of control trees across levels and architectures. \milestone support for the selection of an algorithm and generation of the associated control tree.
\rivet Existing runtimes for controlling parallelism and communication within and across levels
will be integrated and/or extended. % \milestone runtime integration for GPU orchestration. \milestone runtime integration for distributed parallelism.
\end{itemize}

\girder{Support of Legacy and Future Applications}\label{girder:legacy}

A novel, vertically integrated framework brings new opportunities such as breaking out of traditional API layers, exploring families of algorithms (either at runtime or during tuning) to better adapt to specific hardware and software applications, and removing restrictions such as particular data layouts. However, a vast amount of existing infrastructure is already written for existing APIs such as ScaLAPACK. We term these ``legacy'' applications, although of course many such applications are currently cutting-edge and are expected to be used for some time to come. Additionally, static APIs are still useful across many domains where flexibility or new functionality is not required.

\paragraph{Design Goals:}
\begin{itemize}
\item{\bf Supporting legacy applications: } While new scientific software is a driver of innovation and discovery, robust, battle-tested software is critical for sustained and sustainable advancement towards scientific goals. Supporting plug-in compatibility with existing software will further enhance the impact of FAMLIES and expand the user community, %( community metrics listed elsewhere)
 and enable usage instead of, or side-by-side with, LAPACK and its derivatives.   
\item{\bf Enabling innovation and growth: } New scientific software presents an ideal chance to take advantage of new technologies and software development ideas. By moving beyond traditional (and in particular, inflexible or restrictive) APIs, FAMLIES can provide value-added features and functionality which improve usability, performance, and/or scientific impact. %\metric We will assess these  qualities via user surveys and/or case studies.
\item{\bf Flexibility in the definition of interface layers: } The freedom to break out of traditional interface layers not only allows for improved internal design of the framework and exposure of additional optimization opportunities, but also to support new functionality not covered by previous APIs. The flexibility of FAMLIES, provided by the control tree abstraction (Girder~\subref{girder:algorithm}), will also allow for more direct customization of algorithms, as well as techniques such as just-in-time compilation (not planned as part of this proposal, but possible in future work).
\end{itemize}

\paragraph{\bf Project Plans:}

\begin{itemize}
\rivet We will implement traditional LAPACK/LAPACKE interfaces for DLA functionality developed in Girder~\subref{girder:functionality}  will be implemented (\milestone*).
\rivet TAPP interfaces for multi-linear functionality developed in Girder~\subref{girder:functionality}. %will be implemented (\milestone*).
\rivet ScaLAPACK interfaces for distributed DLA functionality developed in Girder~\subref{girder:functionality} will be implemented.%(\milestone*).
\rivet Novel interfaces for DLA and multi-linear algebra functionality which expose additional control over algorithmic and execution details and enable functionality or optimization opportunities beyond traditional APIs will be developed. 
% \metric we will document particular use cases enabled and gather survey feedback.
\end{itemize}

\girder{Representative Functionality}\label{girder:functionality}

While building out the core framework, it will be vitally important to test the integration of all framework components using complete end-to-end functionality, particularly those operations of high importance to the wider community. Implementing representative functionality will provide both basic functionality checks as well as key measures of framework performance. Fully supporting the functionality that is a strict superset of LAPACK and its derivatives is not within scope of this first phase.
%proposal at a level of support available from the CSSI program.

\paragraph{Design Goals:}
\begin{itemize}
\item{\bf Implementing critical ``core'' DLA functionality:} a small number of DLA operations are used by the vast majority of scientific applications, often in a bottle-neck portion of the code. Early detection of   any design or performance issues affecting such operations is critical. We will compare performance for these operations against existing implementations on CPU shared-memory 
%(\metric*), 
GPU,
%(\metric*), a
and distributed-memory architectures (both CPU- and GPU-based).
%\metric*)

\item{\bf Exploring functionality beyond traditional DLA:} While LAPACK incorporates a large number of DLA algorithms (as do ScaLAPACK, MAGMA, etc. to a lesser extent), and new functionality is being added all the time, an even larger space remains unexplored or woefully unoptimized. For example, we recently explored skew-symmetric matrix factorizations, which uncovered large functionality gaps even at the BLAS level.\cite{ltlt_perf} Even more glaring is the lack of functionality for higher-dimensional linear algebra (tensors), which is of mission-critical importance to fields such as computational chemistry and machine learning. Performance for such operations can be measured on both synthetic benchmarks % (\metric*) 
as well as selected operations culled from real-world applications.%(\metric*).
\end{itemize}

\noindent We have pre-identified operations which contribute towards the above design goals. Additional or substitute operations may be identified during the project period based on community feedback.

\paragraph{\bf Project Plans:}

\begin{itemize}
\rivet We will implement linear solvers based on  LU with pivoting,
%(\milestone*),
Cholesky,
% (\milestone*), 
and Householder QR 
%(\milestone*) 
factorizations.

\rivet SVD related functionality will be implemented. % (\milestone*).

\rivet DLA functionality beyond LAPACK/ScaLAPACK 
will be implemented. % \metric Important operations will be identified in collaboration with the community.
\rivet We will implement non-linear tensor factorizations including CP-ALS % (\milestone*)
and HOOI/HOSVD.
%(\milestone*).
\rivet We will implement tensor contraction networks (either pre-ordered or unordered sequences of tensor contractions).
%, \milestone*).
\end{itemize}

\girder{Community}\label{girder:community}

A broad community will contribute to building the framework from conception to application.
We will expand upon the 
 established communities
of the BLIS, libflame, and Elemental projects. 
% See the ``Delivery Mechanism and Community Usage Metrics'' document for additional details.

\paragraph{\bf Design Goals:}
\begin{itemize}
\item 
{\bf Learning from hindsight:} The proposed project builds on  half a century of the theory and practice of high-performance linear and multi-linear algebra library development.  
\item{\bf Expanding a robust community of  advisors, collaborators, users, and contributors:} 
Early and continued participation of a robust community with diverse researchers and users drawn from the scientific community and industry is critical to the short- and long-term success and impact of this initial groundwork project.
\end{itemize}

\paragraph{\bf Project Plans:}
\begin{itemize}
\rivet 
 An advisory committee will be formed from the stakeholders.
 %(\milestone*).  

\rivet 
Insight from experts who were part of previous efforts (such as LAPACK and its derivatives) will be gathered to ascertain what did and did not work regarding the processes of design, implementation, deployment, maintenance, and application. 
% (\milestone*). 

\rivet
The FAMLIES community will be created by  expanding the scope of the active community of developers, contributors, and users that formed around BLIS.
% (\milestone*).

\rivet
Avenues for communication will include GitHub issues and pull requests, a Discord  server, weekly meetings of the core group (frequently attended by external collaborators), 
online meetings of focus groups, the annual retreat,  and a periodic newsletter.
% (\milestone*).
\rivet
 New participants and stakeholders will be attracted and onboarded by using and expanding existing educational materials developed in part by our prior NSF-sponsored projects.  This includes four Massive Open Online Courses (MOOCs) and related notes/activities offered by edX~\cite{edX} (free for auditors), spanning undergraduate and graduate linear algebra~\cite{LAFF,ALAFF-book}, formal derivation \`a la FLAME~\cite{LAFF-On-Correctness}, and the basic ideas underlying BLIS~\cite{LAFF-On-HP}.
These take advantage of the FLAME notation and workflow, thus providing a common language for the theoretical and practical aspects of the project.
% \milestoneA cohort of participants, including PIs, senior personnel, graduate students, undergraduates, and the broader stakeholders, will be onboarded.
\rivet
Efficient procedures, instructions, and mechanisms will be created that reduce the burden on ``gate keepers'' while empowering and encouraging community-provided high-quality contributions to ensure sustainability and inclusiveness.  This will leverage and improve  upon what is in place for BLIS.
% (\milestone*).

\rivet
Self-paced and in-person/hybrid  tutorials will be created.
% (\milestone*). 

\end{itemize}

\section{Impact}
\label{sec:impact}
%\begin{verbatim}
%- Extends FLAME to multi-dimensional tensors
%- Dealing with structures that includes super/sub diagonal
%- 
%\end{verbatim}

% \subsection{Intellectual Merit}

This project will provide fundamental insights into the commonality of data movement,  algorithm description and control, and programming interfaces across diverse functionality (BLAS-like, LAPACK-like, and  tensor computations), levels of parallelism (sequential, shared memory parallel, massively parallel), and architectures (CPU, GPU, and other accelerators). It will also develop and refine techniques for engineering a vertically integrated software framework which can simultaneously achieve often competing goals of readability, maintainability, efficiency, flexibility, extensibility, and usability.
The techniques used to enable all of these advances in a single software framework are significant innovations and intellectual insights into computer science, scientific computing, and how software and hardware interacts when pushing the limits of high performance.

% \subsection{Broader Impacts}
The broader impacts of the project can be divided into multiple categories:
{\bf Broad impact on scientific discovery.} The widespread applicability of the proposed software impacts a broad range of scientific fields. 
    The established collaborations with industry and the national labs will stimulate adoption in commonly-used math libraries and toolsets;
{\bf Education, training, and public outreach.} 
    The abstractions that will be used to vertically integrate the software stack link the theory of numerical algorithms to their practical implementation,  allowing others to use them in innovative ways;
    {\bf Interdisciplinary cooperation.} The project will foster interaction between computer scientists, computational scientists (e.g. chemists, physicists) and industrial partners in order to promote cross-disciplinary solutions and dissemination of idea;
    {\bf Workforce development.} 
    We have a long history of cultivating the careers of people with diverse backgrounds and enabling career changes.    Many former undergraduates and Ph.D. students have found careers in academia and industry.  Our MOOCs have introduced thousands (260,000+ registrations) to fundamental knowledge and the frontiers of the field;
    {\bf Building on existing, recognized capabilities.}
    The eventual production-ready version of the framework will leverage the computing resources provided to NSF grantees, for example through the ACCESS program and at the future LCCF, by potential inclusion in the standard software stack. The BLIS library, developed using previous CSSI funding, is already available at major NSF computing centers such as TACC.

% \subsection{Sustained and Sustainable Impact}

While the proposed framework will have sustained impact by accelerating the pace of scientific discovery in quantum chemistry, in training and deploying large machine learning models, and in other fields, the sustainability of this impact will be ensured through the cultivation of a diverse, highly invested community. In particular, we will continue to forge strong connections with other NSF projects, industry, and the national labs through continuous engagement with collaborators, advisors, and other stakeholders. We also aim to build a stream of industry financial commitments which will contribute to long-term sustainability. This approach has been highly successful with our previous CSSI-funded BLIS project. Educational materials and documentation will be made freely available under open licenses in order to increase the sustainable impact.

\section{Conclusion}

\label{sec:conclusion}

The proposed framework will provide the foundation and framing (girders and rivets) for a modern, vertically integrated dense linear and multilinear algebra software stack.  It is the community that will have to help us finish and  furnish the resulting structure so that it becomes a thriving resource in support of scientific discovery.  
\newpage
\bibliographystyle{unsrt}

\bibliography{biblio,tm}

\end{document}